\def \emr{\mathrm e}

\documentclass[preprint,showpacs,preprintnumbers,amsmath,amssymb]{revtex4}
\usepackage{graphicx}% Include figure files

\begin{document}

\title{A Simple Theory of Condensation}

\author{\firstname{Savely} \surname{Rabinovich}}
\email{shaul@digital-verification.com}
\affiliation{Digital Verification Technology, Israel}

\date{\today}

\begin{abstract}
A simple assumption of an emergence in gas of small atomic
clusters consisting of $c$ particles each, leads to a phase
separation (first order transition).  It reveals itself by an
emergence of ``forbidden'' density range starting at a certain
temperature.  Defining this latter value as the critical
temperature predicts existence of an interval with anomalous heat
capacity behaviour $c_p\propto\Delta T^{-1/c}$.  The value $c=13$
suggested in literature yields the heat capacity exponent
$\alpha=0.077$.
\end{abstract}

\keywords{clustering, first order phase transition, critical point,
specific heat, critical exponent}

\pacs{61.20.Gy, 61.20.Ne}

\maketitle

\smallskip
\section{Introduction}

The theory of gas-liquid condensation is, probably, the most
famous unsolved problem in the classical statistical
mechanics~\cite{Isi73}. Numerous attempts to attack the problem
have been made during the last hundred years.  They were based on
a wide range of different techniques: from cummulant expansion to
field theoretical methods of phase transitions~\cite{Lan00}.  A
considerable step in this direction had been made by the cluster
(droplet) theory of M.E.~Fisher~\cite{Fish67}.  This theory
predicts an essential singularity of the free energy at the
condensation point.

A simple model of condensation which opens the way for appearance
of a critical point and the corresponding phase separation is
suggested here.  This model reveals the basic desirable features
of the condensation and allows a new and self-consistent
definition of the critical point.  Moreover, it identifies the
famous heat capacity singularity and explains it up to the
calculation of the divergency exponent in an excellent accordance
with the measured data.

Isolated clusters of atoms and molecules have been observed
experimentally in molecular beams and studied
theoretically~\cite{Ha94Be99}.  Stability of such clusters has
been studied also in a liquid-like environment by S.~Mossa and
G.~Tarjus~\cite{MoTa03}.  They have shown that the locally
preferred structure of the Lennard-Jones liquid is an icosahedron
(13 atoms), and that the liquid-like environment only slightly
reduces the relative stability of it.

Scattering experiments can also be regarded as an additional
indirect argument in favor of clustering in liquids. For example,
argon radial distribution function~\cite{KRP74KP76} shows neither
temperature nor density dependence of its first maxima abscissae,
i.e. internuclear distances in solid, liquid and gaseous argon are
inherent characteristics of the material.  In other words, this
phase independence can be attributed to the persistence of small
dense clusters.

More detailed study of experimental evidences in favor of the
existence of relatively stable small atomic clusters will be
published elsewhere~\cite{Voronel}.

%%% ----------------------------------------------------------------------

\section{Basic Assumption}

\begin{figure}[!ht]
\includegraphics{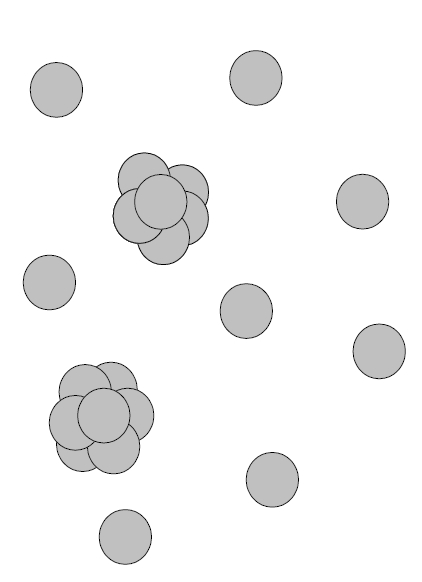}
\caption{\label{fig:fig1}Gas as binary mixture.}
\end{figure}

 Therefore it is possible to formulate the following basic
assumption: {\it elementary particles of gas
(atoms or molecules) form small, relatively stable clusters
consisting of $c$ particles each.} Their concentration is a
function of state. Thus, it immediately infers that the gas should
be considered as (at least) two-component system (see
Fig.~\ref{fig:fig1}).

The ground state of the system under consideration is expected to
be a full separation as the energetically preferable one (we do
not address here those special cases when geometry allows packing
denser that FCC or HCP). On the other hand, at high temperature
the system remains a mixture of atoms and clusters. Thus, at
finite temperature the separation into two phases occurs.

This observation helps us to answer a very natural question: why
do we suppose only one size clusters to be formed or, at least, to
be stable.  Unfortunately, we do not know an {\em a priori} reason
for this.  On the other hand, as one sees, the existence of
clusters of one size leads to the separation.  Therefore, the
existence of clusters of any different number of particles would
reveal itself through multiple separations.  To the best of the
author knowledge, it is not what happens in Nature with simple
liquids. Thus, this {\em a posteriori} argument justifies our
basic assumption. By the way, one may attribute complicated phase
diagrams of complex liquids to the existence of clusters of
different sizes and nature.

Such a model reveals a universal behaviour. Indeed, a close
vicinity of the critical point (if it exists) has to be governed
by the universal properties of the two-component mixture
separation, regardless of the specific details of the
inter-particle interaction. The latter affects the critical
parameters, i.e. physical coordinates, but not the system's
behaviour.

Our basic assumption plays a role analogous to the Cooper pairing
in the superconductivity theories: it is a microscopic phenomenon
underlying the macroscopic one.  Knowledge of the exact (probably,
quantum) mechanism of this clustering is not crucial to understand
the liquid-gas transition.

%%% ----------------------------------------------------------------------

\section{Free Energy}

We start with the expression for the Helmholtz free energy for a
two-component slightly non-ideal gas mixture~\cite{FoGu39}
\begin{eqnarray}\label{eq:FE}
\beta F = N_1 \ln\left( \frac{\lambda_1^3}e \frac{N_1}V \right) +
N_2 \ln\left( \frac{\lambda_2^3}e \frac{N_2}V \right) + \beta E_B
N_1 \nonumber \\
\qquad\qquad +\frac1V \left( B_{11}N_1^2 +2B_{12}N_1N_2
+B_{22}N_2^2 \right).
\end{eqnarray}
Let $N_1=n$ be the number of clusters containing $c$ particles
each, $N_2=N-cn$ and $N$ is the total number of particles.
$\beta=(k_BT)^{-1}$ as usual. As it said, we
assume that all the clusters have the same and constant number of
constituent particles, $c$. $\lambda_i=(2\pi\beta/m_i)^{1/2}\hbar$
is a thermal wave length and $m_i$ is a particle mass. $E_B$
stands for a cluster binding energy and $B_{ij}$ denote second
virial coefficients.  Thus,
\begin{equation}\label{eq:FEn}
\beta F = n\ln\left( \frac{\lambda_1^3}e \frac nV \right) + (N-cn)
\ln\left( \frac{\lambda_2^3}e \frac{N-cn}V \right) + \beta E_B n +
\frac1V B(\beta;n),
\end{equation}
and the internal energy
\begin{equation}\label{eq:IEn}
U = \left(\frac{\partial(\beta F)}{\partial\beta}\right)_V  =
\frac32\frac1\beta [N-(c-1)n] + E_B n + \frac1V B'_\beta(\beta;n),
\end{equation}
where
\begin{equation}\label{eq:Bn}
B(\beta;n) \equiv B_{11}(\beta)n^2 +2B_{12}(\beta)n(N-cn)
+B_{22}(\beta)(N-cn)^2.
\end{equation}
Within the same approximation (slightly non-ideal mixture) the
equation of state reads~\cite{FoGu39}
\begin{equation}\label{eq:EoS}
P\beta = \frac1V[N-(c-1)n] + \frac1{V^2}B(\beta;n).
\end{equation}

A dynamic equilibrium configuration of the two-component system is
defined by the value of $n$ corresponding to the minimum of the
total free energy. Simple differentiation of Eq.~(\ref{eq:FEn})
leads to the main equation determining $n$:
\begin{equation}\label{eq:XE}
\ln\left( \lambda_1^3 \frac nV \right) - c\ln\left( \lambda_2^3
\frac{N-cn}V \right) + \beta E_B + \frac1V B'_n(\beta;n) =0,
\end{equation}
or
\begin{equation}\label{eq:XEf}
\ln\left( \lambda^3 x\rho \right) - c\ln\left( \lambda^3 \rho
(1-cx) \right) - {\textstyle\frac32}\ln c + \beta E_B + \rho
B'_x(\beta;x) =0,
\end{equation}
where $\rho\equiv N/V$, $x\equiv n/N$, $\lambda=\lambda_2$,
$\lambda_1=c^{-1/2}\lambda$ and $B(\beta;x) = B_{11}x^2
+2B_{12}x(1-cx) + B_{22}(1-cx)^2$.

One has to solve analytically Eq.~(\ref{eq:XEf}), i.e. to find
$x=x(\rho)$.  Instead, we found an inverse function,
$\rho=\rho(x)$, where $x\in[0,1/c]$.  It is easily done with the
aid of Lambert $W$-function~\cite{CGHJK96Ha05} (another notation:
$\omega$-function):
\begin{equation}\label{eq:Rho}
\lambda^3 \rho = \left[\frac{ax}{(1-cx)^c}\right]^{\frac1{c-1}}
\exp \left\{ -W\left( - \left[ \frac{ax}{(1-cx)^c}
\right]^{\frac1{c-1}} \frac{B'_x(\beta;x)}{(c-1)\lambda^3}\right)
\right\},
\end{equation}
where $a=c^{-3/2}\exp(\beta E_B)$.  In fact, the equation of
state~(\ref{eq:EoS}) in the form
\begin{equation}\label{eq:EoSrho}
P\beta = \rho[1-(c-1)x] + \rho^2 B(\beta;x)
\end{equation}
and Eq.~(\ref{eq:Rho}) define $P(\rho)$ using the parameter $x$.
The most interesting feature of Eq.~(\ref{eq:Rho}) is the
existence of "forbidden" values for $\rho$.  This behaviour is
governed by the sign of the derivative $B'_x(\beta;x)$. Namely, if
for a given $\beta$ it remains negative for all permissible values
of $x$, then $\rho$ ranges over the entire positive semi-axis.  It
is clear from the behaviour of Lambert function in the negative
range~\cite{CGHJK96Ha05}. If the expression changes its sign to
positive, an equilibrium solution jumps from the $W_0$ branch,
continued from positive argument, to the $W_{-1}$ one. Moreover,
the positive range of the expression will have another "forbidden"
region as the absolute value of the Lambert function's negative
argument cannot exceed $1/\emr$.

%%% ----------------------------------------------------------------------

\section{The Critical Point}

The standard definition of a critical point is
\begin{equation}\label{eq:CrPt}
\left(\frac{\partial P}{\partial\rho}\right)_\beta =
\left(\frac{\partial^2P}{\partial\rho^2}\right)_\beta = 0.
\end{equation}
However, this definition is not applicable if one expects some
singularity to be revealed at this point.  Moreover, as we just
saw, there exists some special behaviour characterized by the sign
of $B'_x(\beta;x)$.  Thus, the very last (critical) point before
the $\rho$-axis becomes ``teared'' up is defined by
$B'_x(\beta_c;x_c)=0$.  In fact, this equation defines critical
parameters: (inverse) critical temperature, $\beta_c$, and
critical concentration, $x_c$, satisfying
\begin{equation}\label{eq:CrEqv}
[B_{12}(\beta_c)-cB_{22}(\beta_c)] +
x_c[B_{11}(\beta_c)-2cB_{12}(\beta_c)+c^2B_{22}(\beta_c)] =0.
\end{equation}
The left-hand side consists of smooth monotonic functions of
$\beta$ (second virial coefficients) and is linear in $x$ and,
thus, attains its extremum at a limiting point. It cannot be
$x_c=0$ because our physical system is supposed to be stable for
small concentrations. Therefore, the only possibility is $x_c=1/c$
and Eq.~(\ref{eq:CrEqv}) becomes
\begin{equation}\label{eq:CrEqvR}
B_{11}(\beta_c)-cB_{12}(\beta_c) =0,
\end{equation}
whose root, $\beta_c$, is the inverse critical temperature.
Naturally, these equations for $x_c$ and  $\beta_c$ are strongly
approximation dependent ones. Higher viral expansion will
complicate Eq.~(\ref{eq:CrEqv}) leading to different values for
the roots $x_c$ and  $\beta_c$.

An important observation to make here is the atom-cluster
($B_{12}$) and the cluster-cluster ($B_{22}$) interactions should
be substantialy weak in comparison with the inter-atomic ($B_{11}$)
one, since part of the gas energy is accumulated in the cluster
bindings. It results, in turn, in ``shallow'' potential well with
a much shorter repulsive part and relatively small inter-cluster
distance and, then, in a much higher density of a heavy component
of the gas.

This new definition of the critical point, $B'_x(\beta_c;x_c)=0$,
allows one to write down an expansion in the vicinity of this point
\begin{equation}\label{eq:CrExp}
B'_x(\beta;x)\approx B''_{x\beta}(\beta_c;x_c)\Delta\beta +
B''_{xx}(\beta_c;x_c)\Delta x,
\end{equation}
where $\Delta\beta \equiv \beta_c-\beta$ and $\Delta x \equiv
x_c-x$.  Substituting this, $x\rightarrow\frac1c$,
$\beta\rightarrow\beta_c$ and $1-cx\rightarrow c\Delta x$ into
Eq.~(\ref{eq:XEf}) we obtain the main Eq.~(\ref{eq:XEf}) in a
close vicinity of the critical point
\begin{equation}\label{eq:XEfCr}
c \ln\left( \Delta x \right) - cA = \rho B''_{xx}\Delta x,
\end{equation}
where $cA\equiv \rho B''_{x\beta}\Delta\beta + \beta_c E_B -
(c-1)\ln\left( \lambda_c^3 \rho\right) + (c-
{\textstyle\frac52})\ln c$.  This equation is solved as before
with the aid of the Lambert function and its solution reads:
\begin{equation}\label{eq:XCrS}
\Delta x = \emr^A \exp\left\{-W\left({-\textstyle\frac1c} \rho
B''_{xx} \emr^A \right)\right\}
\end{equation}
with $\emr^A= \left( \lambda_c^3 \rho\right)^{1-\frac1c}
c^{1-\frac5{2c}} \exp{\left\{ {\textstyle\frac1c} \left( \beta_c
E_B + \rho B''_{x\beta}\Delta\beta \right) \right\} }$.  This
looks like an ultimate solution of the problem, at least, in the
vicinity of the critical point but it does not account for the
basic feature --- the discontinuity of $\rho$-scale --- and it
should be used very carefully.

%%% ----------------------------------------------------------------------

\section{Specific Heat}

The internal energy is given by
\begin{equation}\label{eq:IEx}
\frac UN = \frac32\frac{1-(c-1)x}\beta + E_B x + \rho
B'_\beta(\beta;x)
\end{equation}
and the specific heat --- by
\begin{eqnarray}\label{eq:SH}
c_V = &\frac\partial{\partial T}\left( \frac UN \right)_V = -k_B
\beta^2\frac\partial{\partial\beta}\left( \frac UN \right)_\rho
\nonumber\\ =& k_B \left\{ \textstyle{\frac32}
\left[1-(c-1)x\right] - \rho \beta^2 B''_{\beta\beta} \right\}
  + k_B\beta \left\{ \textstyle{\frac32} (c-1) -
\beta E_B - \rho\beta B''_{x\beta} \right\}x'_\beta.
\end{eqnarray}
Therefore, if one looks for special behaviour of this quantity in
the vicinity of the critical point then $x$ and $x'_\beta$ have to
be examined. We also make use of the fact that $c_V$ on the
critical isohore behaves like $c_p$ in the second order phase
transition.~\cite{LanLif80}

We start with substituting Eq.~(\ref{eq:CrExp}) into
Eq.~(\ref{eq:Rho}) and note that $B''_{xx}(\beta_c;x_c)=
B_{11}(\beta_c)-2cB_{12}(\beta_c)+c^2B_{22}(\beta_c)=
B_{12}(\beta_c)-cB_{22}(\beta_c)$.  It represents the cluster-atom
and the cluster-cluster interactions which are supposed to be very
small. Thus, one can expect existence of an interval where
$B'_x(\beta;x)\approx B''_{x\beta}(\beta_c;x_c)\Delta\beta$ and
\begin{equation}\label{eq:RhoApp}
\ln\left(\lambda_c^3\rho\right) = \ln\left[\frac{a_c/c}{(c\Delta
x)^c}\right]^{\frac1{c-1}} - W\left( - \left[
\frac{a_c/c}{(c\Delta x)^c} \right]^{\frac1{c-1}}
\frac{B''_{x\beta}(\beta_c;x_c)}{(c-1)\lambda_c^3} \Delta\beta
\right).
\end{equation}

Further consideration depends on the sign of $B''_{x\beta}
(\beta_c;x_c) \Delta\beta$.  In the homogeneous phase it is
negative and we are on the $W_0$ branch with a small positive
argument.  Here it is enough to take~\cite{CGHJK96Ha05}
$W_0(y)\approx y$ and subsequently
\[
\lambda_c^3\rho = \left[\frac{a_c/c}{(c\Delta x)^c} \right]
^{\frac1{c-1}} \left\{1 - \left[ \frac{a_c/c}{(c\Delta x)^c}
\right]^{\frac1{c-1}} \frac{B''_{x\beta}(\beta_c;x_c)}
{(c-1)\lambda_c^3} \Delta\beta \right\}.
\]
The relevant root behaves as
\[
\left[\frac{(c\Delta x)^c}{a_c/c} \right]^{\frac1{c-1}} \approx
\frac{B''_{x\beta}(\beta_c;x_c)} {(c-1)\lambda_c^3} \Delta\beta
\qquad\mbox{or}\qquad \Delta x \propto
\left(\Delta\beta\right)^{1-\frac1c}.
\]
It means that the derivative $\Delta x/\Delta\beta$ and therefore
the specific heat will show here the famous dependence $c_p\propto
\left(\Delta\beta\right)^{-\frac1c}$.  In view of the previous
suggestion, $c=13$, this exponent becomes $\alpha\approx0.077$.

An analogous calculation cannot be done for a nonhomogeneous
phase as an equilibrium solution does not exist in this region.

%%% ----------------------------------------------------------------------

\section{Conclusions}

A model that explains basic features of the condensation is
presented.  A simple assumption of relative stability of only one
type of clusters statistically emerging in the gas immediately
leads to the first order phase transition (phase separation) 
at some finite temperature.  It is
experimentally observed as a condensation process.

It should be stressed that this model is by no means a simplified
version of Fisher's one.  As much as the monogamy is not a
simplified version of the polygamy and the monotheism is not
a simplified version of the polytheism.

Mathematically, the condensation reveals itself as a forbidden
density (volume) region. The density jumps from its gaseous value
to the liquid one.  No intermediate values are allowed.  A
correspondent region for the Van der Waals equation is a
well-known S-shaped instability.  It needs special auxiliary
construction to be treated as a metastable state.

This paper presents a new concept of the critical point: it is a
point of the density's continuity failure.  This definition
coincides graphically with the old one but it allows to construct
a convenient expansion in the close vicinity of the point under
consideration.  It demonstrates the famous singularity with the
exponent $\alpha\approx0.077$ that agrees excellently with
known data.

%%% ----------------------------------------------------------------------

\section*{Acknowledgement}

The author is grateful to Alexander Voronel and Moshe Schwartz for
the valuable discussions. A financial support of Alexander Voronel
during a part of this study is kindly acknowledged.  Extensive
editorial efforts of Ely Klepfish made this manuscript readable.

\end{document}